\documentclass[preprint,showpacs,prl,preprintnumbers,amsmath,amssymb]{revtex4}

\usepackage{graphicx}
\usepackage{dcolumn}
\usepackage{bm}

\begin{document}
\title{non-Local Quantum Criticality in Ce(Ru$_{1-x}$Fe$_x$)$_2$Ge$_2$ ($x$=$x_c$=0.76)}

\author{W. Montfrooij$^{1)}$, M. C. Aronson$^{1)}$, B. D. Rainford$^{2)}$, J. A. Mydosh$^{3)}$, A.P. Murani$^{4)}$, P. Haen$^{5)}$, and T. Fukuhara$^{6)}$}

\address{$^{1)}$ University of Michigan, Ann Arbor, MI 48109, USA\\
$^{2)}$ University of Southampton, Southampton SO17 1BJ, United Kingdom\\
$^{3)}$ Leiden University, 2300 RA Leiden, the Netherlands\\
$^{4)}$  Institute Laue-Langevin, F-38042, Genoble Cedex 9, France\\
$^{5)}$  CRTBT-Grenoble, BP 166, F-38061, Grenoble, France\\
$^{6)}$ Faculty of Engineering, Toyama Prefectural University, Toyama 939-0398, Japan.}

\begin{abstract}
{We present the results of inelastic neutron scattering
measurements, performed near the antiferromagnetic quantum
critical point in Ce(Ru$_{0.24}$Fe$_{0.76}$)$_2$Ge$_2$. Both local
and long range fluctuations of the local moments are observed, but
due to the Kondo effect only the latter are critical. We propose a
phenomenological expression which fits the energy $E$, temperature
$T$ and wave vector $q$ dependences of the generalized
susceptibility $\chi(q,E,T)$, describing the non-Fermi liquid
$E/T$ scaling found at every wave vector.}
\end{abstract}
\maketitle


The stability of magnetic order among moments exchange coupled to
the conduction electrons in a metal arises from the dynamic
competition of two forces: the Kondo compensation of the moments,
and long range intermoment coupling driven by the RKKY
interaction, both depending on the strength of the exchange
interaction $J$. An extensive body of experimental evidence has
been compiled\cite{stewart,sachev} which shows that magnetic order
is initially stabilized by increasing $J$, but that ultimately
magnetic order terminates at a zero temperature quantum critical
point (QCP) for a critical value $J_{c}$, yielding a magnetically
enhanced but paramagnetic Fermi liquid phase
for larger values of $J$.\\
Two different scenarios have been proposed to explain this generic
phase diagram. In the first view\cite{hertz,millis,coleman2},
magnetic order at the QCP is a spin density wave (SDW) instability of the Fermi
surface, as it is preceded by moment compensation below a Kondo
temperature $T_K$ which is finite at $J$=$J_{c}$. As for a
conventional second order phase transition occurring at finite
temperature, only the long wavelength fluctuations of the order
parameter are  critical in this scenario, leading to Lorentzian
energy and wave vector dependences in the vicinity of the ordering
wave vector $\vec{q}^*$\cite{hertz,millis}. In the second
view\cite{coleman2,schroeder,si}, known as the local moment
scenario, both $T_K$ and the magnetic ordering temperature are
zero at the QCP. Complete Kondo compensation of moments is thus
only possible in the paramagnetic phase for $J>J_{c}$. Magnetic
order, which occurs for $J<J_{c}$, consequently involves moments
which are both long-lived and spatially localized. Here, the
competition between Kondo screening and long range order leads to
local as well as long wavelength degrees of freedom, and both can
be critical near the QCP. The experimental signatures of the local
moment scenario are an anomalous energy dependence for all wave
vectors \cite{si}, as well as the absence of Kondo compensation at
any temperature scale when $J$=$J_{c}$.\\
Despite the very different roles for moment compensation in these
two theoretical scenarios, it is in practice difficult to
distinguish them, since experiments are necessarily performed over
a range of finite temperatures and on real materials with disorder
and compositional inhomogeneities. Neutron scattering experiments
have been central to this explication, although to date they have
only been carried out on two systems with $T$=0 phase transitions,
CeCu$_{6-x}$Au$_x$($x$=$x_{c}$=0.1)\cite{schroeder} and
UCu$_{5-x}$Pd$_x$($x$=1,1.5) \cite{meigan2,meigan}. In both
materials, strong scattering in the experimental energy window
implies the presence of localized magnetic moments, while the
modulation of the static susceptibility $\chi_{q}$ reveals the
presence of residual magnetic interactions among these moments.
Different arguments are presented for the two materials which
support the view that the local degrees of freedom and not the
long range coupling control the critical behavior. Direct
comparison of $\chi_{q}$ at different $T$ in UCu$_{5-x}$Pd$_{x}$
($x$=1,1.5) \cite{meigan2} shows that these interactions are
almost unaffected by lowered temperature, while $\chi_q$ measured
at the largest $q$, representing the local susceptibility,
diverges as $T\rightarrow$0. Excitations at every $q$ display
anomalous energy dependences as well as $E/T$ scaling.\\
A different depiction of quantum critical behavior is found in
CeCu$_{5.9}$Au$_{0.1}$, where the commensurate roles of
temperature, energy, and wave vector in tuning the dynamic susceptibility $\chi(q,E,T)$ to
criticality are reflected in a generalized Curie-Weiss
expression\cite{schroeder}, $\chi(q,E,T)^{-1} \sim
\theta_{q}^{\alpha}+(iaE+T)^{\alpha}$. While the longest lived and
longest ranged fluctuations occur for $\vec{q}$=$\vec{q}^{*}$, the
propagation wave vector of the parent antiferromagnetic (AF) phase
where the Weiss field $\theta_{q}$ vanishes, anomalous $E$ and $T$
dependences are found at every wave vector, as well
as $E/T$ scaling when $\theta_{q=q^*}$=0. A dynamical mean-field
model has been developed \cite{si} which argues that the $T$=0
phase transition in CeCu$_{5.9}$Au$_{0.1}$ is actually driven by
fluctuations on the shortest length scales, providing an
operational definition of local
criticality.\\
We present here the results of a neutron scattering study on a new
system, Ce(Ru$_{0.24}$Fe$_{0.76}$)$_2$Ge$_2$, which has been doped
to an AF QCP. Localized moments are also present in this system,
although they experience substantial Kondo compensation. Both
local and long wave length fluctuations are found at all
temperatures, although only the latter are truly critical.
Accordingly, we argue that Ce(Ru$_{0.24}$Fe$_{0.76}$)$_2$Ge$_2$ is
the first example of a quantum critical system in which the $T$=0
instability is SDW-like. non-Fermi liquid (nFl) $E/T$ scaling is
found at all temperatures and wave vectors, but neither the
generalized Curie-Weiss expression\cite{schroeder} nor the
Lorentzians of the mean field SDW model\cite{millis} describe our
data. We propose instead a new phenomenological expression for
$\chi(q,E,T)$ which is only critical on the longest length scales.\\
The phase diagram obtained from both pressurization \cite{sullow}
and Fe doping studies of CeRu$_{2}$Ge$_{2}$ (Fig. \ref{cordata}a)
fulfills the basic requirement for the SDW scenario, displaying an
AF QCP accompanied by a finite Kondo temperature.
CeRu$_{2}$Ge$_{2}$ is a ferromagnet (FM)
\cite{rietschel,brian,fontes} with a Curie temperature of 8 K.
Under pressure, ferromagnetism is supplanted by two AF phases,
which in turn vanish at a critical pressure of 67 kbar.
Substantial moment compensation is observed at the critical
pressure, where $T_K$ is estimated to be $\sim$ 15
K. Further increase of the pressure stabilizes a Fermi liquid
phase over an expanding range of temperatures. Inelastic neutron
scattering experiments cannot be performed at these high
pressures, so it is significant that a similar sequence of phases
is observed in the doping series
Ce(Ru$_{1-x}$Fe$_{x}$)$_{2}$Ge$_2$ (see Fig. \ref{cordata}a) and
for CeRu$_{2}$(Si$_{1-x}$,Ge$_{x}$)$_{2}$ \cite{sullow}. The FM
phase boundary is only qualitatively reproduced by the resistivity
data \cite{fontes}, but the AF phase boundary, taken from our ac
and dc susceptibility measurements reproduces the pressure
results. The AF QCP occurs for a critical Fe concentration
$x_{c}$=0.76$\pm$0.05 \cite{nofontes}.\\
We prepared a 30 g polycrystalline sample of
Ce(Ru$_{0.24}$Fe$_{0.76}$)$_2$Ge$_2$ for neutron scattering
experiments by arc-melting, followed by a two week anneal at 1000
$^{o}$C. Microprobe measurements verified that the intended composition
varies by only a few percent across a representative piece cut
from the neutron scattering sample. Isolated regions of an
impurity phase, CeGe$_2$, were also found, occupying no more than
2\% of the sample volume. Neutron diffraction experiments found AF
correlations below 5 K, although no bulk phase transition was
observed down to 1.7 K\cite{wouter}. We conclude that our sample
is very close to the
$x_c$= 0.76 QCP.\\
Neutron scattering experiments were carried out
at the IN6 time-of-flight spectrometer at the
Institute Laue-Langevin, using an incident neutron wavelength
of 5.12 \AA. The data were corrected for
self-attenuation and neutron absorption, and the magnetic
components were separated from the non-magnetic components (Bragg
peaks, phonon and multiphonon contributions, and nuclear
incoherent scattering) using a non-magnetic reference sample,
LaRu$_2$Ge$_2$. Fig. \ref{cordata}b shows the absolutely
normalized dynamic structure factor $S(q,E)$ obtained from this
process for $T$= 7.5 K.\\
The strong scattering found at all $q$ in Fig.
\ref{cordata}b attests to the intrinsically localized character
of the fluctuating moments in
Ce(Ru$_{0.24}$Fe$_{0.76}$)$_2$Ge$_2$. A quantitative measure of
this localized moment is obtained by direct integration of
$S(q,E,T)$ over the energies accessed in this experiment, yielding
a fluctuating moment of 0.8 $\mu_{B}$ at 20 K. Since this is a
substantial fraction of the 1.74 $\mu_{B}$ expected for the crystal field split groundstate of the Ce$^{3+}$
ions\cite{brian}, we conclude that these moments are
long lived on the time scales of our experiment, with a
substantial degree of spatial localization at all $T$.\\
The scattering is strongly enhanced at the smallest $q$,
revealing the presence of long-ranged interactions among these
moments at low temperature. In order to analyze these
interactions, we extract $\chi_{q}$
using the Kramers-Kronig relation:
\begin{equation}
\chi_q=\int^{\infty}_{-\infty}\frac{\chi^{''}(q,E)dE}{\pi E}=
\int^{\infty}_{-\infty}\frac{S(q,E)(1-e^{-E/k_BT})dE}{\pi E},
\label{kramers}
\end{equation}
where $k_B$ is Boltzmann's constant. We can expand the range of
energies available from the neutron scattering experiment, and
improve the accuracy of the Kramers-Kronig transformation by first
fitting the measured $\chi^{''}(q,E)$ to
a modified Lorentzian line shape, described below.
The static susceptibility $\chi_{q}$ deduced from this analysis is
plotted at several temperatures in Fig. \ref{cordata}c, revealing
two striking features, also evident in the corrected $S(q,E)$ data
shown in Fig. \ref{cordata}b. First, for all $T$ and $q>$
1\AA$^{-1}$, $\chi_q$ is virtually independent of $q$. As we will
show below, the line shape as a function of $E$ is independent of
$q$ in this q-range. Both observations indicate that this
q-independent susceptibility describes the local response of
spatially localized magnetic moments to fluctuating magnetic
fields. $\chi_q$ also shows a pronounced enhancement at small $q$,
which increases with decreasing $T$.  We attribute this to the
growth of critical correlations associated with incipient
AF-order, at the small and incommensurate propagation vector
$\sim$0.2$\pm $0.1 \AA$^{-1}$ implied by the onset of AF-order at
satellite positions ($q$= 1.56, 1.74 \AA$^{-1}$) around the (101)
Bragg reflection ($q$= 1.64 \AA$^{-1}$), found in our neutron
diffraction
experiments \cite{wouter}.\\
$\chi_{q}$ shows clear signs of moment
compensation by the Kondo effect at the lowest $T$. The
temperature dependences of $\chi_q(T)$ for representative q-values
are plotted in Fig. \ref{bunch}a, including the uniform ($q$= 0)
susceptibility $\chi_{o}(T)$, obtained from a dc susceptibility
measurement\cite{dc} on a 35 mg piece taken from the neutron
scattering sample. We find that $\chi_{o}(T)$ increases with
decreasing temperature, displaying the hallmark power law behavior
for $T<$ 20 K found in many nFl systems: $\chi_0(T) =
C_0/(T+\theta_{W})^{\alpha}$. Here, the Weiss-temperature
$\theta_W$= 0.9 $\pm$ 0.2 K, $C_0$=0.87$\pm$ 0.03
$\mu_B^2$/meV$^{0.49}$ and $\alpha$ = 0.51 $\pm$ 0.01. Since
$\theta_{W}$ is non zero, we see that $\chi_{o}(T)$ does not truly
diverge as $T$$\rightarrow$ 0.  The absence of divergence is even
more evident in the local susceptibility which is
found by integrating $\chi_q$ over the experimental q-range, $\chi_{loc}(T)$ =$\int^{q_{max}}_0 \chi_q(T)d\vec{q}/(4/3\pi q_{max}^3)$.
$\chi_{loc}$, which is indistinguishable from $\chi_q$ for $q>$1\AA$^{-1}$, saturates for $T$$<$ 5 K (see Fig \ref{bunch}a).
Further evidence\cite{bickers} for partial Kondo compensation
comes from the $\sim$ 25$\%$ reduction in scattered intensity as
the temperature is lowered from 20 K to 1.8 K (see Fig.
\ref{bunch}a), despite a
simultaneous narrowing of the energy line width.\\
The dynamic response in Ce(Ru$_{0.24}$Fe$_{0.76}$)$_2$Ge$_2$ is
dramatically different from those found in fluctuating moment
systems far from quantum critical points. Like previous neutron
scattering studies on Kondo lattice systems
\cite{brian,loewenhaupt,aeppli,goldman,broholm,walter}, the
dynamic response in Ce(Ru$_{0.24}$Fe$_{0.76}$)$_2$Ge$_2$ is broad
and quasi-elastic. However, Fig. \ref{bunch}b shows that the
Lorentzian line shape \cite{brian,loewenhaupt,aeppli,goldman,broholm,walter}  common to Kondo lattices agrees very poorly
with our measured $\chi(q,E)$, both for a small wave number,
0.35\AA$^{-1}$ where the moments are interacting, and at a large
wave number 1.15 \AA$^{-1}$ where the response is purely local. We
find instead that at all $E$, for all $T$ from 1.9 K - 200 K, and
for the complete q-range probed in our experiment that our data
can be satisfactorily described by a simple phenomenological
expression:
$\chi(q,E)=\chi_q(T)/[1-iE/\Gamma_q(T)]^{\beta}$.\\
The observed line shape $\chi$(q,E) is controlled by an energy
scale $\Gamma_q(T)$, and by a dynamical exponent $\beta$ = 0.15
$\pm$ 0.05. The temperature dependence of $\Gamma_q(T)$ which we
extract from these fits is plotted for two values of $q$ in Fig.
\ref{bunch}c. For $q$=1.15 \AA$^{-1}$, $\Gamma$(q) has the
familiar T-dependence of a Kondo impurity system,
initially decreasing with $T$ before saturating and increasing
weakly below $T_K$, which we identify as $\sim$ 5
K. Dynamics on this local length scale are consequently not
critical. In contrast, $\Gamma_q$ determined for $q$= 0.35
\AA$^{-1}$ approaches zero with decreasing $T$, as expected for
critical slowing-down associated with the $T$= 0 phase transition.
With the exception of the lowest $T$ at the largest $q$,
$\Gamma_q$ is approximately linear in $T$,
$\Gamma_q$=$\theta_q+a_{q}T$. The q-dependence of $\theta_q$ is
shown in Fig. \ref{bunch}d, demonstrating that dynamic
criticality, i.e. $\Gamma_q \rightarrow$ 0, can only be achieved
if $\theta_q \rightarrow$0. For
Ce(Ru$_{0.24}$Fe$_{0.76}$)$_2$Ge$_2$ this occurs as $q$ approaches
the propagation wave vector of incipient AF order, 0.2$\pm$0.1
\AA$^{-1}$. We conclude that at all temperatures $\chi(q,E,T)$ is
dominated at short length scales by the excitations of individual
Kondo moments, while the long-wave length fluctuations become
increasingly long-lived and
ultimately critical as T$\rightarrow$ 0.\\
A random phase approximation (RPA) analysis of $\chi_q(T)$ shows
that the singular behavior found near the QCP in
Ce(Ru$_{0.24}$Fe$_{0.76}$)$_2$Ge$_2$  requires the collaboration
of both local and long-range dynamical correlations. The
long-range correlations are graphically demonstrated in Fig. 3a,
where we have plotted $\chi_{loc}(T)$/$\chi_{q}(T)$ at several
temperatures. An increasing suppression of $\chi_{loc}$/$\chi_{q}$
is observed at small $q$ as the temperature is lowered, signalling
the growth of long range AF coupling. An estimate of this
coupling, $U_q(T)$ can be obtained in the RPA approximation by
noting that $\chi_{loc}(T)/\chi_q(T)= 1-U_q(T)\chi_{loc}(T)$.
$U_{q}$ deduced from this analysis at 4.4 K is plotted in the
inset of Fig. 3a, demonstrating that the interactions are long
ranged, rapidly vanishing for wave vectors larger than $\sim$ 0.6
\AA$^{-1}$. As shown in Fig. 3b, the temperature dependences of
$U_q$ are very different for large and small $q$.
$U$($q$=0.55\AA$^{-1}$) is almost temperature independent, while
$U$($q$=0.275\AA$^{-1}$) increases almost a factor of four between
20 K and 1.5 K. In contrast, $\chi_{loc}(T)$ initially increases with
decreasing temperature, but ultimately saturates below $\sim$5 K.
This RPA analysis reveals that local fluctuations initially
provide a bias towards criticality in
Ce(Ru$_{0.24}$Fe$_{0.76}$)$_2$Ge$_2$, but ultimately it is the
intermoment coupling $U_q(T)$ which actually drives criticality. A
similar conclusion was reached for U$_{2}$Zn$_{17}$\cite{broholm},
a heavy fermion antiferromagnet with a N\'{e}el temperature $T_N$=
9.7 K. However, neither a departure from Lorentzian line shape,
nor any other nFl effects
were observed in this system, which is not at a QCP.\\
The modified Lorentzian introduced above and the observed
T-linearity of $\Gamma_q(T)$ implies that our data should
also display nFl $E/(T+\theta_q)$ scaling. To confirm this, we
have plotted $\chi^{''}(q,E)(T+\theta_W)^{0.51}$  for $q$= 0.35,
0.55, 1.35 and 1.75\AA$^{-1}$ as functions of
$E$/k$_{B}$($T+\theta_q$) in Fig. 4. An excellent collapse of the
data taken at different temperatures is observed at each $q$,
spanning three orders of magnitude in the scaling variable
$E$/k$_{B}$($T+\theta_q$). The exponent 0.51 is taken from the
temperature dependence of $\chi_0(T)$, as was also found in the
locally critical systems UCu$_4$Pd\cite{meigan2,meigan} and
CeCu$_{5.9}$Au$_{0.1}$\cite{schroeder}. Unlike those systems,
we find in Ce(Ru$_{0.24}$Fe$_{0.76}$)$_2$Ge$_2$ that the
dynamical scaling function itself requires a second exponent $\beta$= 0.15.\\
Our neutron scattering measurements have established that there
are local moments present in
Ce(Ru$_{0.24}$Fe$_{0.76}$)$_{2}$Ge$_{2}$  which experience
increasing AF coupling as $T\rightarrow$0. Although fluctuations
of these moments are observed on every length scale, $\chi_q$ only
diverges at the residual ordering wave vector of the nearby AF
phase. Correspondingly, we observe a substantial Kondo suppression
of the local fluctuations below $\sim$ 20 K. The dominance of the
long wave length correlations at the lowest temperatures, and the
finite Kondo temperature at the QCP together imply that the $T$=0
antiferromagnetic transition in
Ce(Ru$_{0.24}$Fe$_{0.76}$)$_{2}$Ge$_{2}$ is a collective
instability of the strongly interacting quasiparticles, and is not
locally critical. The mean field view of such a phase transition
requires a single diverging length scale with an accompanying
diverging time scale, leading to Lorentzian lineshapes for $\chi$
in energy and wave vector. We find instead that the susceptibility
$\chi(q,E,T)$ is well described by a modified Lorentzian
expression, encompassing the $E/T$ scaling which we observe at
every wave vector.

We acknowledge stimulating discussions with P. Coleman, A. J.
Millis, and Q. M. Si. MCA thanks T. Gortenmulder and R. Hendrikx
for invaluable technical assistance, and acknowledges the
hospitality of the MSM group at Leiden during the early stages of
this project. We thank I. P. Swainson for carrying out the
neutron diffraction measurements. Work at the University of
Michigan was supported by NSF-DMR-997300.

\begin{figure*}[b]
\caption[]{(a): The magnetic phase diagram for CeRu$_{2}$Ge$_{2}$
as functions of pressure \cite{sullow}(dashed lines) and Fe doping
($\diamond$: FM phase boundary \cite{fontes}, $\star$:
AF phase boundary, present work). Filled circles
represent the pressure dependent Kondo temperature, which is
finite at the QCP \cite{sullow}. The solid lines are guides to the eye. The electrical
resistivity is quadratic in temperature \cite{sullow} in the shaded part of the
phase diagram (FL). (b): The fully corrected $S(q,E)$ for
Ce(Ru$_{0.24}$Fe$_0.76$)$_2$Ge$_2$ as a function of neutron
momentum transfer $\hbar$$q$ and energy transfer $E$, for $T$= 7.5
K. (c): The static susceptibility $\chi_q$ for $T$= 2.9 K
($\diamond$), 7.5 K ($\star$) and 15.2 K ($\bigtriangleup$). Note
the incipient AF order around the (101) Bragg peak (q=1.64
\AA$^{-1}$) at 2.9 K.} \label{cordata}
\end{figure*}
\begin{figure*}[b]
\caption[]{(a): $\chi_q(T)$ for $q$= 0.35\AA$^{-1}$($\diamond$),
for $q$= 0.45\AA$^{-1}$($\bigtriangleup$) and integrated over all
$q$, $\chi_{loc}(T)$ ($\star$). The solid line is the $q$ = 0, dc
susceptibility $\chi_{0}$. Also shown ($\bullet$, in $\mu_B^2$) is the average
of $S_q(T)$ over
 1 $< q <$ 1.8 \AA$^{-1}$, demonstrating the onset of Kondo-screening at
 $T\approx$ 20 K.
 (b): $\chi"(q,E)/E$ for $q$= 0.35 \AA$^{-1}$ ($\bullet$) and $q$= 1.15 \AA$^{-1}$ ($\circ$) at $T$= 4.4 K.
 The data
 at $q$= 1.15 \AA$^{-1}$ have been divided by 2 for the sake of plotting clarity. The solid lines are the best
 fits to
 the modified Lorentzian line shape described in the text,  with $\beta$= 0.15. The dashed curve is the best fit
 Lorentzian line shape ($\beta$= 1).
(c): The energy linewidth $\Gamma_{q}$ of $\chi"(q,E)/E$ for $q$=
0.35 \AA$^{-1}$ ($\bullet$) and $q$= 1.15 \AA$^{-1}$ ($\circ$).
(d): q-dependence of the residual linewidth $\theta_q$, explained
in the text. Also shown is $\theta_W$ at $q$=0.} \label{bunch}
\end{figure*} 
\begin{figure*}[b]
\caption[]{(a): $\chi_{loc}(T)$/$\chi_{q}(T)$ as a function of $q$
for $T$=1.9 K ($\diamond$), 4.4 K ($\bigtriangleup$), 7.5 K
($\square$), and 15.2 K ($\star$). As explained in the text, this
quantity is directly related to the interaction $U_q$, which is
plotted in the inset for T=4.4 K. (b): While the temperature
divergence of $\chi_{loc}$ ($\star$) is cut off below $\sim$ 5 K by
the Kondo effect, $U$($q$=0.275\AA$^{-1}$) ($\bullet$) increases
monotonically to the lowest $T$. For $q$=0.55 \AA$^{-1}$
($\diamond$), $U_q$ is T-independent.} \label{j}
\end{figure*}
\begin{figure*}[b]
\caption[]{Scaling of the dynamic response for various q-values.
The neutron scattering data $\chi"(q,E)$ have been multiplied by
$(T+\theta_{W})^{\alpha}$ ($\alpha$= 0.51, see text), and
displayed versus the reduced variable $E/(T+\theta_q)$, with
$\theta_q$ as in Fig. \ref{bunch}d. Note that the various q-values
are offset by half a decade along the vertical axis. The
temperatures range from 1.9 K (darkest symbols) to 200 K (lightest
symbols), and there is substantial overlap in $E/(T+\theta_{q})$
among the 11 temperatures displayed in this figure. Every scaling
curve represents about 3000 independent data points.}
\label{scaling}
\end{figure*}


\begin{thebibliography}{10}
\bibitem{stewart} G. R. Stewart, Rev. of Mod. Physics {\bf 73}, 797 (2001).

\bibitem{sachev} S. Sachdev, {\it Quantum Phase Transitions} (Cambridge University Press, Cambridge,
England, 1999).
\bibitem{hertz} J.A. Hertz, Phys. Rev. B $\bf{14}$, 1165 (1976).
\bibitem{millis} A.J. Millis, Phys. Rev. {\bf B48}, 7183 (1993); private communication.
\bibitem{coleman2} P. Coleman, Physica {\bf B259-261}, 353 (1999).
\bibitem{schroeder} A. Schr\"{o}der {\it et al.},
Nature {\bf 407}, 351 (2000).
\bibitem{si} Qimiao Si {\it et al.}, Nature {\bf 413}, 804 (2001).
\bibitem{meigan2} M.C. Aronson {\it et al.},
Phys. Rev. Lett. {\bf 75}, 725 (1995).
\bibitem{meigan} M.C. Aronson {\it et al.},
Phys. Rev. Lett. {\bf 87}, 197205 (2001).
\bibitem{sullow} S. S\"{u}llow {\it et al.}, Phys. Rev. Lett. $\bf{82}$,
2963 (1999).
\bibitem{rietschel} H. Rietschel {\it et al.}, J. Magn. Magn. Mater. {\bf 76-77}, 105 (1988); A. Boehm {\it et al.}, J. Magn. Magn. Mater. {\bf 76-77}, 150 (1988)
\bibitem{fontes} M.B. Fontes {\it et al.}, Phys. Rev. {\bf B53}, 11678 (1996).
\bibitem{brian} B.D. Rainford and S.J. Dakin, Phil. Mag. {\bf B65}, 1357 (1992).
\bibitem{nofontes} This value is below the estimate of Fontes {\it et al.}\cite{fontes},
however, their results are not inconsistent with $x_c$= 0.76.
\bibitem{wouter} W. Montfrooij {\it et al.}, unpublished.
\bibitem{dc} $\chi_{o}$ in Ce(Ru$_{0.24}$Fe$_{0.76}$)$_{2}$Ge$_{2}$ has substantial anisotropy, reflecting preferred orientation and/or
Ising-like behavior. However, we only use the T-dependence of $\chi_0$ in our analysis, not
its absolute value.
\bibitem{bickers} N.E. Bickers, D.L. Fox and J.W. Wilkins, Phys. Rev. {\bf B36}, 2036 (1987).
\bibitem{loewenhaupt} M. Loewenhaupt {\it et al.}, Journal de Physique {\bf 40}, C4-142 (1979).
\bibitem{aeppli} G. Aeppli, E. Bucher and G. Shirane, Phys. Rev. {\bf B32}, 7579 (1985).
\bibitem{goldman} A.I. Goldman {\it et al.}, Phys. Rev. {\bf B33}, 1627 (1986).
\bibitem{broholm} C. Broholm {\it et al.}, Phys. Rev. Lett. {\bf 58}, 917 (1987).
\bibitem{walter} U. Walter, M. Loewenhaupt, E. Holland-Moritz and W. Schlabitz, Phys. Rev. {\bf B36}, 1981 (1987).
\end{thebibliography}
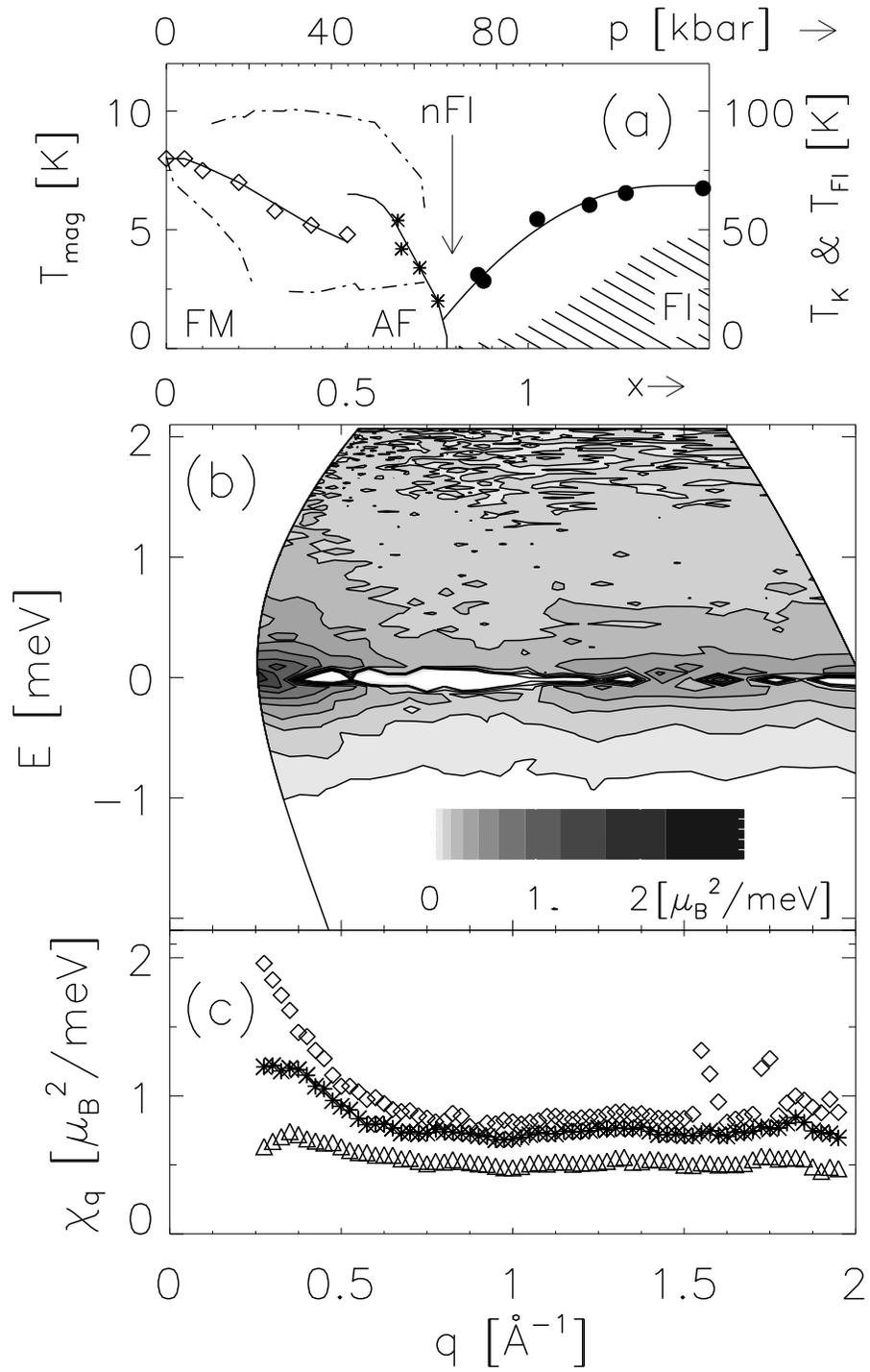
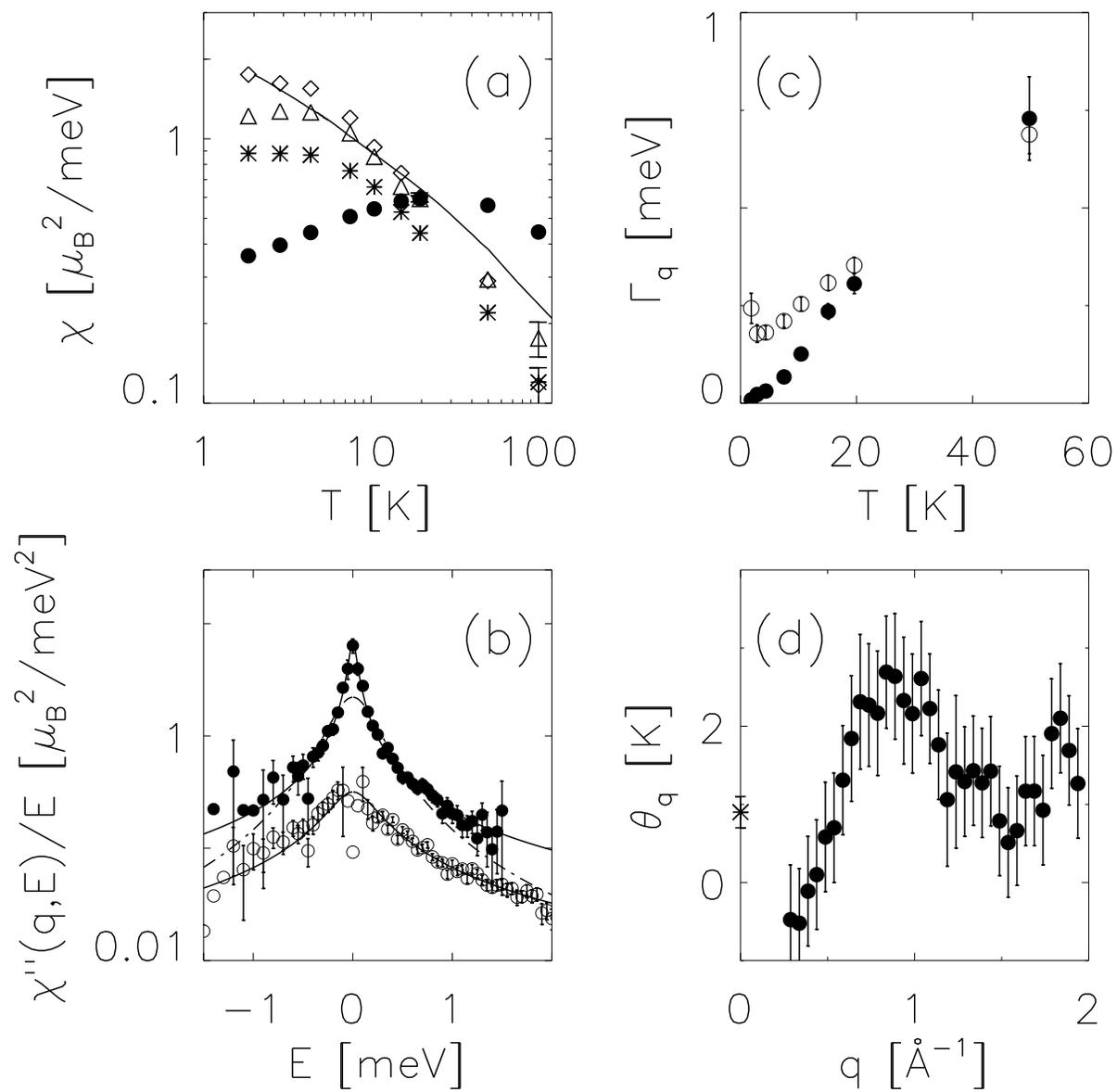
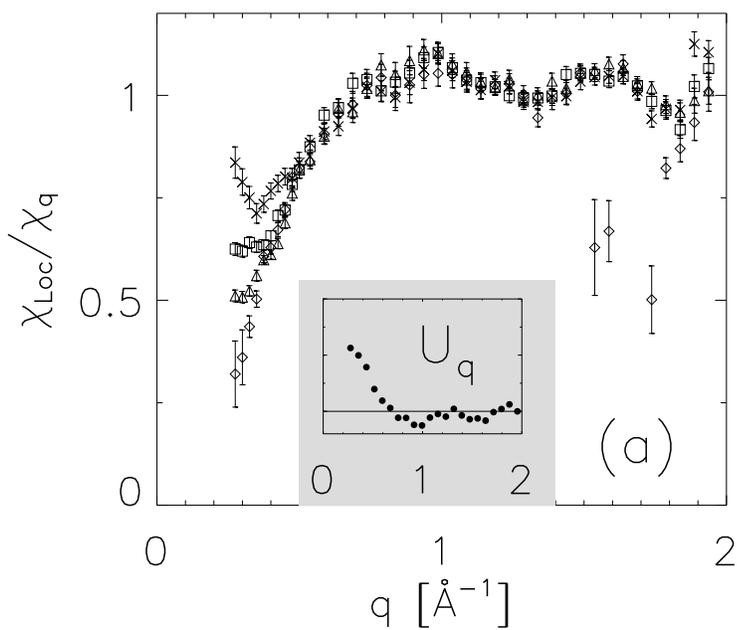
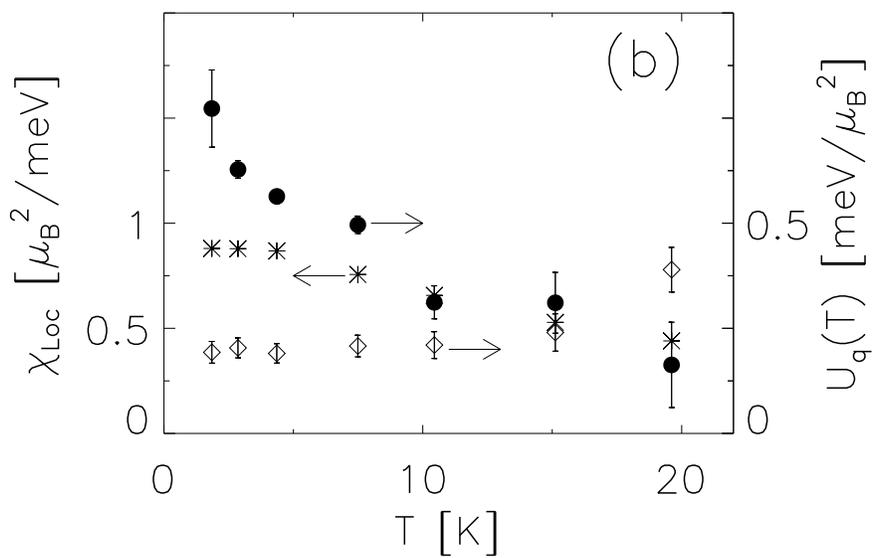
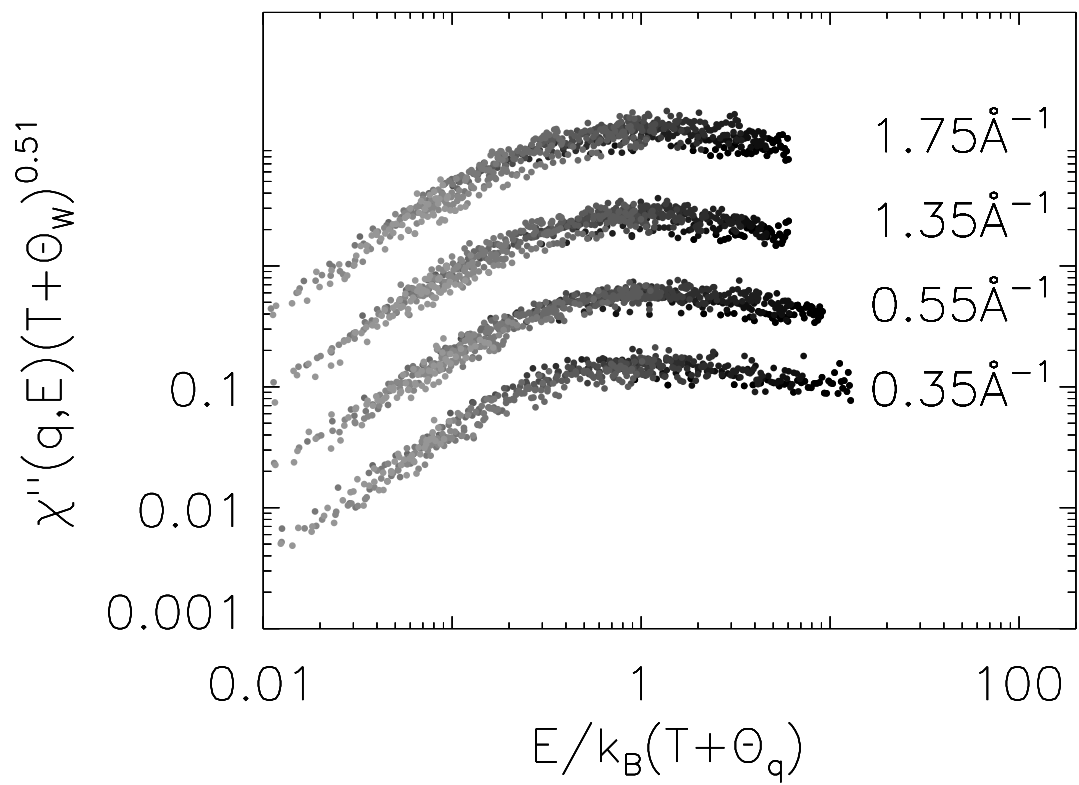
\end{document}